\documentclass[12pt]{iopart}

\begin{document}
\newcommand{\Kbp}{\bar{K}^{\prime}}
\newcommand{\w}{\omega}
\newcommand{\D}{\Delta}
\newcommand{\Om}{\Omega}
\newcommand{\la}{\lambda}
\newcommand{\lab}{\bar{\lambda}}
\newcommand{\La}{\Lambda}
\newcommand{\ep}{\epsilon}
\newcommand{\beq}{\begin{equation}}
\newcommand{\eeq}{\end{equation}}
\newcommand{\bal}{\begin{align}}
\newcommand{\eal}{\end{align}}
\newcommand{\al}{\alpha}
\newcommand{\alb}{\bar{\alpha}}
\newcommand{\alp}{\alpha^{\prime}}
\newcommand{\albp}{\bar{\alpha^{\prime}}}
\newcommand{\n}{\eta}
\newcommand{\nb}{\bar{\eta}}
\newcommand{\np}{\eta^{\prime}}
\newcommand{\nbp}{\bar{\eta^{\prime}}}
\newcommand{\psib}{\bar{\psi}}
\newcommand{\psip}{\psi^{\prime}}
\newcommand{\psibp}{\bar{\psi^{\prime}}}
\newcommand{\phib}{\bar{\phi}}
\newcommand{\phip}{\phi^{\prime}}
\newcommand{\phibp}{\bar{\phi^{\prime}}}
\newcommand{\z}{z}
\newcommand{\zb}{\bar{z}}
\newcommand{\zp}{z^{\prime}}
\newcommand{\zbp}{\bar{z^{\prime}}}
\newcommand{\f}{f}
\newcommand{\fb}{\bar{f}}
\newcommand{\fp}{f^{\prime}}
\newcommand{\fbp}{\bar{f^{\prime}}}
\newcommand{\g}{g}
\newcommand{\gb}{\bar{g}}
\newcommand{\gp}{g^{\prime}}
\newcommand{\gbp}{\bar{g^{\prime}}}
\newcommand{\definedas}{:=}

\title{Discrete path integral for spin-$\frac{1}{2}$ in the Grassmannian representation}

\author{S Shresta}
\address{Department of Physics, University of Maryland, College Park, MD 20742, USA}
\begin{abstract}
A variation to the usual formulation of Grassmann representation path integrals is presented. Time-indexed anticommuting partners are introduced for each Grassmann coherent state variable and a general method for handling the effect of these introduced Grassmann partners is also developed. These Grassmann partners carry the nilpotency and anticommutivity of the Grassmann coherent state variables into the propagator and allow the propagator to be written as a path integral. Two examples are introduced in which this variation is shown to yield exact results. In particular, exact results are demonstratated for a spin$-\frac{1}{2}$ in an time-dependent magnetic field, and for a spin-boson system. The stationary path approximation is then shown to be exact for each example.
\end{abstract}


\section{Introduction}

In the use of coherent state path integrals to describe spin systems three approaches have been prevalent. One approach has been to describe the spin degrees of freedom via coherent states of SU(2) or on the sphere\cite{alscher1,alscher2}. Another approach is to use a boson mapping or a stereograhic projection of the sphere onto the complex plane\cite{boudjeda}. A third approach, which is the one taken here, is to use Grassmannian coherent states to represent the spin. One drawback of this approach is that it is restricted to representations of spin-$\frac{1}{2}$ or two-level systems. Within the Grassmannian representation there are also two variations. One approach is to generate Grassmannian coherent states by an exponentation of the single fermion creation operators. The other approach is to generate them by exponentiation of spin increasing operators. The advantage of the first approach is that the Hamiltonian so defined always has definite even parity since the fermion operators always appear in pairs, thereby avoiding nonlinear terms in the action due to the mixed Grassmann parity of the Hamiltonian. As has been pointed out, this fact has caused the first approach to be predominant\cite{smirnov}. However an advantage to the second approach is that the interaction terms in many Hamiltonians of interest remain bilinear. Thereby allowing more straightforward evaluation of the propagator. When the advantages of these two approaches are combined consistently an altered formulation of the Grassmann coherent state path integral emerges which is more useful.

The further contribution that this work hopes to make to the field is simply to suggest an addition to the formulation of Grassmann coherent state path integrals that help make that framework more useful and consistent. The addition is the introduction of a time-indexed anticommuting partner to each Grassmann variable, which carries the nilpotency and anticommutivity of the Grassmann variables into the propagator. It makes the Grassmann representation more useful and consistent in that it extends its range of validity beyond previous formulations\cite{anastopoulos}, offers alternative expansions, and provides a simpler route to non-Markovian reduced dynamics in that representation. In the first section the framework is sketched roughly. In the subsequent sections two simple examples are developed and shown to yield exact results. The two examples are the one of a spin$-\frac{1}{2}$ in a time dependent B-field, and of a spin$-\frac{1}{2}$ interacting with a single quantized mode of the EM field (the Jaynes-Cummings Hamiltonian). Both cases are taken for an infinitely heavy particle (i.e. motional degrees of freedom are ignored). For the two simple examples shown, each is analysed exactly and then demonstrated to also be exact under the stationary phase approximation. 

Grassmannian coherent states were first formulated for use in a path integral by Ohnuki and Kashiwa \cite{ohnuki}. An excellent review of their properties is available from Cahill and Glauber \cite{cahill}. Those of the bosonic coherent states are detailed by Perelomov \cite{perelomov}. Some of the most important properties are the following. The bosonic and fermionic coherent states are constructed by exponentiation of their respective creation operators acting on their respective reference vacuum states. Due to the algebra which their creation and annihilation operators satisfy, the coherent states are then eigenstates of their respective annihilation operators.
\begin{equation}
\begin{array}{lr}
 | \z \rangle = {\rm e}^{a^\dagger z} | 0 \rangle & a | \z \rangle = \z | \z \rangle \\
| \n \rangle = {\rm e}^{S_+ \n} | \downarrow \rangle & S_- | \n \rangle =\n | \n \rangle  
\end{array}
\end{equation}
The coherent states are an overcomplete set of states with a resolution of unity and measure.  
\begin{equation}
\begin{array}{lr}
1 = \int {\rm d}\mu(z) | \z \rangle \langle \zb | & {\rm d}\mu(\z) = d\z d\zb {\rm e}^{-\zb \z} \\
1 = \int {\rm d}\mu(\n) | \n \rangle \langle \nb | & {\rm d}\mu(\n) = d\n d\nb {\rm e}^{-\nb \n} 
\end{array}
\end{equation}
Note that these Grassmann coherent states are not single fermion coherent states that are generated by the fermionic creation operator. As a result, Hamiltonians which contain single spin up/down operator terms will contain odd terms in this representation.

\section{Current approach}

The approach taken in this paper follows the standard path integral approach with a few differences. In particular, the guiding principle has been to evaluate all quantities discretely as far as possible. The continous limit is used only for the final expressions. It begins by writing the propagator in a coherent state representation and breaking it into a product of discrete time infinitesmal propagators,
\begin{eqnarray} \fl
K(t,0) = \langle \alb_f | U(t,0) | \al_i \rangle  \nonumber\\
\lo{=}  \int \prod_{j=1}^{N-1} {\rm d}\mu(\al_j) \langle \alb_f | U_{N,N-1} | \al_{N-1} \rangle \langle \alb_{N-1} | U_{N-1,N-2} | \al_{N-2} \rangle ... \langle \alb_1 | U_{1,0} | \al_0 \rangle ,
\end{eqnarray}
where $\al$ is a generic symbol for a set of even and odd coherent state variables.

The boundary condition applied is $\al_0 = \al_i$ and $\alb_N = \alb_f$. This choice of boundary condition corresponds to the perspective of unitarily evolving the initial(final) ket(bra) vector and then taking the inner product of the evolved vector with the final(initial) bra(ket) vector. Therefore no additional terms in the action are needed to impose boundary conditions. It is interesting to note that for an orthogonal representation the inner product between the final state and the evolved initial state is zero unless the initial state evolves exactly to the final state. The distinction between final states and evolved initial states is thus unneccessary. 

At this point the product of infinitesmal propagators can not be naturally combined into a single exponential, as is desirable in a path integral formulation. The reason is that there can be odd terms which anticommute in the infinitesmal propagators. With bosonic path integrals that is not a problem since c-numbers commute. In order to avoid this problem a time-indexed anticommuting partner is introduced to all fermionic coherent state variables. Then the propagator becomes a discrete time path integral of a single exponential,
\begin{equation} 
\label{dtcspi1} K(t,0) =\int \prod_{j=1}^N {\rm d}\mu(\al_j) \exp( \alb_f \al_N - \frac{{\rm i}\ep}{\hbar} \sum_{i=1}^N H_{i,i-1} ).
\end{equation}
Although the introduction of the Grassmann partners clearly allow the propagator to be written as above in \eref{dtcspi1} (since they make each infinitesmal propagator even), that does not justify their introduction or elucidate their use. The justification for introducing the anticommuting partners is that they are a counting tool that helps to preserve the trucations and signs of formal expressions. This is most clearly illustrated in a recursive evaluation of the propagator. For example, the propagator for an infinitesmal interval is
\begin{equation}
K(\ep,0) = \exp( \alb_1 \al_0 - \frac{{\rm i}\ep}{\hbar} H_{1,0} ) = \exp(\alb_1 \psi_1 +\phi_1)
\end{equation}
with $\psi_1$ and $\phi_1$ being Hamiltonian dependent and containing a mixture of even and odd terms. For two infinitesmal intervals the propagator is
\begin{eqnarray} \fl
K(2\ep,0) = \int {\rm d}\mu(\al_1) \exp(\alb_2 \al_1 - \frac{{\rm i}\ep}{\hbar} H_{2,1}) K(\ep,0) \nonumber\\ 
\lo{=} \int {\rm d}\mu(\al_1) \exp(\psib_2 \al_1 +\phib_2) \exp(\alb_1 \psi_1 +\phi_1) \nonumber\\
\lo{\neq} \int {\rm d}\mu(\al_1) \exp(\psib_2 \al_1 +\alb_1 \psi_1 +\phib_2 +\phi_1) .
\end{eqnarray}
The last inequality is due to the odd parts of the exponents. However, if one introduces anticommuting partners to the Grassmann coherent state variables then the inequality becomes an equality when done with standard Grassmann integration techniques. The anticommuting partners change the exponents above from Grassmann odd to Grassmann even. A key point is that if anticommuting partners were not introduced to the Grassmann coherent state variables then nonlinear terms would appear in the exponent\cite{smirnov}. The distinction may seem small, but one must at later points respect the anticommutation properties of the Grassmann partners. More explicit examples are given later. 

A recursive evaluation which continues the above single step to succesive infinitesmal unitary evolutions can now be performed. After each evolution anticommuting Grassmann partners are introduced at that time index and the propagator is rewritten in a standard form to facilitate the next evolution,
\begin{eqnarray}
\fl K(\ep,0) = \exp(\alb_1 \al_0 - \frac{{\rm i}\ep}{\hbar} H_{1,0}) = \exp(\alb_1 \psi_1 +\phi_1) \\
\fl K(2\ep,0) = \int {\rm d}\mu(\al_1) \exp(\alb_2 \al_1 - \frac{{\rm i}\ep}{\hbar} H_{2,1}) K(\ep,0) = \exp(\alb_2 \psi_2 +\phi_2) \\
. \nonumber\\
. \nonumber
\end{eqnarray}
By continuing this process one finds the full propagator to be,
\begin{eqnarray}
\fl K(N\ep,0) = \int {\rm d}\mu(\al_{N-1}) \exp(\alb_N \al_{N-1} - \frac{{\rm i}\ep}{\hbar} H_{N,N-1}) K((N-1)\ep,0) \nonumber\\
\lo{=} \exp(\alb_N \psi_N +\phi_N) .
\end{eqnarray}
At each step new exponents are defined from the previous ones,
\begin{eqnarray}
\psi_j = f( \psi_{j-1}, \phi_{j-1} )\\
\phi_j = g( \psi_{j-1}, \phi_{j-1} ) ,
\end{eqnarray}
with $f$ and $g$ some functions specific to the Hamiltonian. This method explicitly gives only exact expressions.  

The propagator in this form is only formally valid because the introduction of the time-indexed anticommuting parts to the couplings causes various truncations in the polynomial expansion that must be respected. The next step is thus to expand the propagator in a polynomial series 
\begin{equation} 
\label{genprop1} K(t,0) = \exp(\alb_f \psi_N +\phi_N) = \sum_{m=0}^\infty \frac{1}{m!} (\alb_f \psi_N +\phi_N)^m ,
\end{equation}
and use the finite difference equations for the exponents and the anticommutation properties to find finite difference equations for the terms in the polynomial series. 
\begin{eqnarray} 
\label{geneom1} [\psi^m]_j = [f(\psi_{j-1},\phi_{j-1})]^m  \\ 
\label{geneom2} [\phi^m]_j = [g(\psi_{j-1},\phi_{j-1})]^m  \\ 
\label{geneom3} [\psi^m]_j [\phi^n]_j = [f(\psi_{j-1},\phi_{j-1})]^m [g(\psi_{j-1},\phi_{j-1})]^n .
\end{eqnarray}
These new equations can then be taken to the continous limit and solved. Finally, putting the solutions back into the expansion of the propagator and resumming gives the propagator. Only in the form of \eref{genprop1} with \eref{geneom1}-\eref{geneom3} does the propagator cease to be a formal expression and allow explicit calculation.

\section{Examples}

\subsection{Spin-$\frac{1}{2}$ in a general time-dependent classical magnetic field} 

To illustrate the use of time-indexed anticommuting couplings the following is a calculation of the evolution of a spin-$\frac{1}{2}$ in a classical magnetic field. The simplest non-trivial case is that of a spin in a $B_z$ field with the addition of a possibly time-dependent $B_x$ and $B_y$ field. The Hamiltonian for this system is
\begin{equation}
\fl H =\gamma \vec{S} \cdot \vec{B} = \frac{1}{2} \hbar \w S_z + \hbar B_x S_x + \hbar B_y S_y = \hbar \w S_+ S_- -\frac{1}{2} \hbar \w + \hbar (S_+ B + B^* S_-).
\end{equation}
Here it is written in a ``hermitian'' form in anticipation of the addition of a Grassmann part to the classical field. The propagator between initial and final Grassmann coherent states is
\begin{equation}
K(t,0) = \langle \nb_t  | \exp[-\frac{{\rm i}}{\hbar} \int_0^t H(s) ds] | \n_0 \rangle .
\end{equation}
In the usual way ($t=N \ep$) the propagator can be time sliced into a discrete time formulation. The propagator for one infinitesmal time step is (up to $O(\ep)$ and dropping the constant term)
\begin{equation} 
\label{1stepprop1} \fl K(j,j-1)=\langle \nb_j | \exp(-\frac{{\rm i}}{\hbar} H \ep) | \n_{j-1} \rangle = \exp[(1-{\rm i} \w \ep) \nb_j \n_{j-1} -\nb_j ({\rm i} B_j \ep) - ({\rm i} B_j^* \ep) \n_{j-1}].
\end{equation}
With eqn(\ref{1stepprop1}) the propagator for a single infinitesmal step can be written down,
\begin{equation}
\fl K(\ep,0) = \exp[(1-{\rm i} \w \ep) \nb_1 \n_0 -\nb_1 ({\rm i} B_1 \ep) - ({\rm i} B_1^* \ep) \n_0] = \exp(\nb_1 \n_1 -\phi_1)
\end{equation}
and the propagator for two infinitesmal time steps is
\begin{eqnarray}
\fl K(2 \ep,0) =\int {\rm d}\mu (\n_1) \langle \nb_2 | \exp(-\frac{{\rm i}}{\hbar} H_2 \ep) | \n_1 \rangle \langle \nb_1 | \exp(-\frac{{\rm i}}{\hbar} H_1 \ep) | \n_0 \rangle \nonumber\\
\lo{=} \exp[(1-{\rm i} \w \ep)^2 \nb_2 \n_0 ] [ 1  -\n_2 ({\rm i} B_2 \ep) - ({\rm i} B_1^* \ep) \n_0 - B_2 B_1^* \ep^2 \nb_2 \n_0 - B_2^* B_1 \ep^2 \nonumber\\
-\nb_2 ({\rm i} B_1 \ep) (1-{\rm i} \w \ep) -(1-{\rm i} \w \ep) ({\rm i} B_2^* \ep) \n_0 + B_2 B_1^* \ep^2 (1-{\rm i} \w \ep)^2 \nb_2 \n_0 ].
\end{eqnarray}
These two propagators have very different forms. However if at this point a time-indexed anticommuting part is given to the classical field such that $\{ B_n, B_m \} = 0$ and $\{ B_n, \n \} =\{ B_n, \nb \} =0$ then the $2 \ep$ propagator can be rewritten as a single exponential,
\begin{eqnarray}
\fl K(2\ep,0)= \exp\{\nb_2 [-{\rm i} B_2 \ep + (1-{\rm i} \w \ep)(-{\rm i} B_1 \ep + (1-{\rm i} \w \ep) \n_0)] \nonumber\\
+ [ -{\rm i} B_1^* \ep \n_0 - ({\rm i} B_2^* \ep)(-{\rm i} B_1 \ep  +(1-{\rm i} \w \ep) \n_0)]\} \nonumber\\
\lo{=} \exp(\nb_2 \n_2 + \phi_2)
\end{eqnarray}
with the definitions
\begin{eqnarray}
\n_2 = (1-{\rm i} \w \ep) \n_1 -{\rm i} B_2 \ep \\
\phi_2 = \phi_1 - {\rm i} B_2^* \ep \n_1.
\end{eqnarray}
Now the 2$\ep$ propagator is in the same form as the $\ep$ propagator. This facilitates a recursive evaluation, so that process can be continued to find the propagator for any number of steps,
\begin{equation}
K(j\ep,0) = \exp(\nb_j \n_j + \phi_j)
\end{equation}
with the recursive definitions
\begin{equation} 
\label{ex1eom1} 
\begin{array}{ll}
\n_j = (1-{\rm i} \w \ep) \n_{j-1} -{\rm i} B_j \ep & \n_0 = \n_i \\
\phi_j = \phi_{j-1} - {\rm i} B_j^* \ep \n_{j-1}  & \phi_0 = 0 .
\end{array}
\end{equation}
Inserting the boundary condition $\nb_N=\nb_f$, one gets for the full propagator
\begin{equation}
K(t=N\ep,0)=\exp(\nb_f \n_N +\phi_N),
\end{equation}
with the variables $\n_N$ and $\phi_N$ defined by eq(\ref{ex1eom1}).

The propagator in the above form can not yet be shown to satify the Schroedinger equation because it hides a major pitfall. The pitfall is that it is a formal expression and has meaning only as a polynomial expansion. Due to the introduction of the time-indexed anticommuting part in the magnetic field, many terms in the polynomial expansion truncate due to the nilpotency of the Grassmann variables. However this is not a weakness, but a strength, since the truncation of polynomial expansions is the reason Grassmann variables were introduced. If the continous limit were taken at this point the correct expansion of the exponential propagator would be lost. Expanding the propagator gives
\begin{equation} 
\label{bprop} K(t,0) = \exp(\nb_f \n_N +\phi_N) = \sum_{m=0}^\infty \frac{[\phi^m]_N}{m!} (1+ \nb_f \n_N).
\end{equation}
In the expansion above the Grassmann variable $\nb_f$ causes a truncation. Analogously, in the $m^{th}$ order terms such as $[\phi^m]_N$, the time-indexed Grassmann parts of the magnetic field cause a truncation. That is, $\phi_N$ is a sum of terms containing many products of Grassmann variables. Products of these coefficients have many terms that are truncated due to nilpotency of the Grassmann variables. Keeping track of the truncations in the final coefficients would be a formidable task, however doing so in the infinitesmal equations of motion is sufficient. For example, instead of calculating $[\phi^m]_N$ by calculating $\phi_N$ first, one can find a differential equation for $[\phi^m]_N$ and calculate it directly. The functions that need to be calculated are thus $[\phi^m]_N$ and $[\phi^m\n]_N$. Adhering to the anticommutation rules one finds (up to $O(\ep)$),
\begin{eqnarray} 
\label{bexeom1} [\phi^m]_j = [\phi^m]_{j-1} - {\rm i} m B_j^* \ep [\phi^{m-1} \n]_{j-1} \\ 
\label{bexeom2} [\phi^m \n]_j = (1-{\rm i} \w\ep)[\phi^{m} \n]_{j-1} - {\rm i} B_j \ep [\phi^m]_{j-1}.
\end{eqnarray}
The above equations can now safely be taken to the continous limit,
\begin{eqnarray}
\label{bconteom1} \frac{{\rm d}}{{\rm d}t}[\phi^m]_t &= - {\rm i} m B^*(t) [\phi^{m-1} \n]_t \\ 
\label{bconteom2} \frac{{\rm d}}{{\rm d}t}[\phi^m \n]_t &= -{\rm i} \w [\phi^m \n]_t - {\rm i} B(t) [\phi^m]_t,
\end{eqnarray}
and used to show that the propagator satisfies the Schroedinger equation (see Appendix A). The propagator \eref{bprop} and \eref{bconteom1}-\eref{bconteom2} give a novel expansion of the propagator and equations for the terms in its expansion. The Schroedinger equation can be reformed from it, but in the expanded form it may be possible to apply new approximations. This issue is addressed in future work.

Having introduced and justified the introduction of the Grassmann partners, they can now be used to rewrite the propagator as a true path integral. The propagator for finite time is
\begin{eqnarray}
\fl K(t,0) =\int \prod_{j=1}^{N-1} {\rm d}\mu (\n_j) \langle \nb_N | \exp(-\frac{{\rm i}}{\hbar} H \ep) | \n_{N-1} \rangle \langle \nb_{N-1} | \exp(-\frac{{\rm i}}{\hbar} H \ep) | \n_{N-2} \rangle ... \nonumber\\
\times \langle \nb_{2} | \exp(-\frac{{\rm i}}{\hbar} H \ep) | \n_{1} \rangle \langle \nb_{1} | \exp(-\frac{{\rm i}}{\hbar} H \ep) | \n_{0} \rangle.
\end{eqnarray}
Due to the anti-commuting properties of the Grassmann variables, the infinitesmal propagators in the above expression could not be combined into a single exponential if a time-indexed anticommuting part were not introduced. After their introduction the propagator becomes,
\begin{eqnarray}
\fl K(t,0)=\int \prod_{j=1}^{N-1} {\rm d}^2 \n_j \mbox{ } \exp \{\nb_N \n_N \nonumber\\
+\sum_{j=1}^N \left[ -\nb_j \n_j + (1-{\rm i} \w \ep) \nb_j \n_{j-1} -{\rm i} \nb_j B_j \ep -{\rm i} B_j^* \ep \n_{j-1} \right] \}.
\end{eqnarray}
One may now evaluate this discrete path integral at the saddle point. Varying discretely, the discrete equation for the stationary path is found to be
\begin{equation} 
\label{beom1} \n_j = (1-{\rm i} \w\ep) \n_{j-1} -{\rm i} B_j \ep
\end{equation}
and the propagator is
\begin{equation}
K(t,0)=\exp \{\nb_N \n_N +\sum_{j=1}^N \left[-({\rm i} B_j^* \ep) \n_{j-1} \right] \}.
\end{equation}
Or, defining again the variable
\begin{equation} 
\label{beom2} \phi_j = -{\rm i} \sum_{i=1}^j B_i^* \ep \n_{i-1} = \phi_{j-1} -{\rm i} B_j^* \ep \n_{j-1},
\end{equation}
and inserting the correct boundary conditions $\nb_N=\nb_f$ and $\n_0=\n_i$, one gets for the propagator
\begin{equation}
K(t,0)=\exp(\nb_f \n_t +\phi_t),
\end{equation}
with the variables $\n_N$ and $\phi_N$ defined by \eref{beom1} and \eref{beom2}. This is the same as the exact result previously derived. This example was handled, in the stationary path approximation, using a boson mapping in \cite{boudjeda} and using the SU(2) representation in \cite{alscher1}. The result found  here of exactness of the stationary path approximation agrees with the same result found in those references.

\subsection{Spin-Boson} 

The Jaynes-Cummings Hamiltonian for a spin$-\frac{1}{2}$ interacting with a single em field mode is
\begin{equation}
H = \hbar \w_o S_+ S_- + \hbar \w a^{\dagger} a + \hbar (S_+ \la a + a^{\dagger}\la S_-).
\end{equation}
Here again it is written in a ``hermitian'' form in anticipation of the addition of a Grassmann part to the spin-boson coupling constant. The propagator between initial and final coherent states is
\begin{equation}
K(t,0) = \langle \nb_t \zb_t | \exp[-\frac{{\rm i}}{\hbar} \int_0^t H(s) ds] | \n_0 z_0 \rangle .
\end{equation}
In the usual way ($t=N \ep$) the propagator can be time sliced into a discrete time formulation. The propagator for one infinitesmal time step is (up to $O(\ep)$)
\begin{eqnarray}
\fl K(j,j-1)=\langle \nb_j \zb_j | \exp(-\frac{{\rm i}}{\hbar} H \ep) | \n_{j-1} z_{j-1} \rangle \nonumber\\
\lo{=} \exp[(1-{\rm i} \w \ep) \zb_j \z_{j-1} +(1-{\rm i} \w_o \ep) \nb_j \n_{j-1} -\nb_j ({\rm i} \la_j \ep) z_{j-1} - \zb_j ({\rm i} \la_j \ep) \n_{j-1}].
\end{eqnarray}
Using this equation the single infinitesmal step propagator is,
\begin{equation}
K(\epsilon,0) = \langle \nb_1 \zb_1 | U_{\epsilon} | \n_0 \z_0 \rangle = \exp[\nb_1 (\psi_1 +\g_1) +\zb_1 (\f_1 +\phi_1)]
\end{equation}
with the definitions,
\begin{equation}
\begin{array}{ll}
\g_1 = ({\rm i} \la_2 \ep) \z_0 & \psi_1 = (1-{\rm i} \w_o \ep) \n_0  \\
\f_1 = (1-{\rm i} \w \ep) \z_0 & \phi_1 = ({\rm i} \la_2 \ep) \n_0
\end{array}
\end{equation}
The 2$\ep$ propagator is then computed from the above to be,
\begin{equation} 
\label{onemode} \fl K(2 \epsilon,0) = \langle \nb_2 \zb_2 | U_{2 \epsilon} | \n_0 \z_0 \rangle = \int {\rm d}\mu (z_1) {\rm d}\mu (\n_1) \langle \nb_2 \zb_2 | U_{\epsilon} | \n_1 \z_1 \rangle \langle \nb_1 \zb_1 | U_{\epsilon} | \n_0 \z_0 \rangle 
\end{equation}
which gives the following unwieldy expression,
\begin{eqnarray}
\fl K(2 \epsilon,0)= \exp[(1-{\rm i} \omega_o \ep) \nb_2 \psi_1 + (1- {\rm i} \w \ep) \zb_2 f_1] \nonumber\\
\lo{\times} \bigg [ 1+ {\rm i} \la \ep \zb_2 \psi_1 +(1-{\rm i} \omega_o \ep) \nb_2 g_1 + {\rm i} \la \ep \nb_2 f_1 + (1- {\rm i} \w \ep) \zb_2 \phi_1 + {\rm i} \la \ep \zb_2 \g_1\nonumber\\
+ {\rm i} \la \ep \nb_2 \phi_1 + {\rm i} \la \ep \nb_2 \phi_1 (1- {\rm i} \w \ep) \zb_2 f_1 - {\rm i} \la \ep \zb_2 g_1 (1-{\rm i} \w_o \ep) \nb_2 \psi_1 \bigg ]
\end{eqnarray}
At this point a time-indexed anticommuting part is given to the coupling constants such that $\{ \la_n, \la_m \} = 0$ and $\{ \la_n, \n \} =\{ \la_n, \nb \} =0$. The $2 \ep$ propagator can be rewritten as a single exponential in the same form as the $\ep$ propagator,
\begin{equation}
K(2 \epsilon,0) = \exp[\nb_2 (\psi_2 + g_2) + \zb_2 (\phi_2 +f_2)]
\end{equation}
with the definitions,
\begin{equation} 
\label{ex2eom1}
\begin{array}{ll}
g_2 = (1-{\rm i} \w_o \ep) g_1 + ({\rm i} \la_2 \ep) f_1 & \psi_2 = (1-{\rm i} \w_o \ep) \psi_1 + ({\rm i} \la_2 \ep) \phi_1 \\
f_2 = ({\rm i} \la_2 \ep) g_1 + (1-{\rm i} \w \ep) f_1 & \phi_2 = ({\rm i} \la_2 \ep) \psi_1 + (1-{\rm i} \w \ep) \phi_1
\end{array}.
\end{equation}
Or for greater ease of use,
\begin{equation}
K(2 \epsilon,0) = \exp(\nb_2 \n_2 + \zb_2 \z_2)
\end{equation}
with the definitions,
\begin{eqnarray} 
\n_2 = (1-{\rm i} \w_o \ep) \n_1 + ({\rm i} \la_2 \ep) \z_1 \\ 
\z_2 = ({\rm i} \la_2 \ep) \n_1 + (1-{\rm i} \w \ep) \z_1 .
\end{eqnarray}
This process can be continued to find the propagator for any number of infinitesmal steps, with the result,
\begin{equation}
K(j \epsilon,0) = \exp(\nb_j \n_j + \zb_j \z_j)
\end{equation}
and the definitions,
\begin{eqnarray} 
\label{ex2eom2} \n_j &= (1-{\rm i} \w_o \ep) \n_{j-1} + ({\rm i} \la_j \ep) \z_{j-1} \\
\label{ex2eom3} \z_j &= ({\rm i} \la_j \ep) \n_{j-1} + (1-{\rm i} \w \ep) \z_{j-1} .
\end{eqnarray}
After inserting the correct boundary conditions $\nb_N=\nb_f$, $\n_0=\n_i$, $\zb_N=\zb_f$, and $\z_0=\z_i$ the propagator for time $t=N\ep$ is
\begin{equation}
K(t=N \epsilon,0) = \exp(\nb_f \n_N + \zb_f \z_N)
\end{equation}
with the variables $\n_N$, and $\z_N$ defined by \eref{ex2eom2} and \eref{ex2eom3}.

As in the previous example the propagator in the above form is only a formal expression and has meaning only as a polynomial expansion. Many terms in the polynomial expansion truncate due to the nilpotency of the Grassmann variables. Expanding the propagator gives
\begin{equation} 
\label{sbprop} K(t,0) = \exp(\nb_f \n_N + \zb_f \z_N) = \sum_{m=0}^\infty \frac{(\zb_f)^m [\z^m]}{m!} (1 +\nb_f \n_N).
\end{equation}
As before differential equations are found for the functions in the expansion of the propagator. The functions that need to be calculated are $[\z^m]_N$ and $[\z^m\n]_N$. Adhering to the anticommutation rules one finds (up to $O(\ep)$),
\begin{eqnarray}
[\z^m]_j = (1 -{\rm i} m\w\ep)[\z^m]_{j-1} - {\rm i} m \la_j \ep [\z^{m-1}\n]_{j-1} \\
\big [\z^m\n]_j = (1-{\rm i} m\w\ep-{\rm i} \w_o \ep)[\z^{m}\n]_{j-1} - {\rm i} \la_j \ep [\z^{m+1}]_{j-1}.
\end{eqnarray}
Or in the continous limit,
\begin{eqnarray} 
\label{sbconteom1} \frac{{\rm d}}{{\rm d}t}[\z^m]_t = -{\rm i} m\w [\z^m]_t - {\rm i} m \la [\z^{m-1} \n]_t \\ 
\label{sbconteom2} \frac{{\rm d}}{{\rm d}t}[\z^{m}\n]_t = (-{\rm i} m\w -{\rm i} \w_o)[\z^{m}\n]_t - {\rm i} \la [\z^{m+1}]_t.
\end{eqnarray}
The propagator of \eref{sbprop} can now be shown to satisfy the Schroedinger equation (see Appendix B). As in the previous example the propagator \eref{sbprop} and \eref{sbconteom1}-\eref{sbconteom2} give a novel expansion of the propagator and equations for the terms in its expansion. However, in this case the unexpanded expression may offer an advantage when seeking the reduced dynamics. In that case the final state of the e.g. boson can be traced out using the formal exponential version of \eref{sbprop}, leaving a formal expression for the reduced propagator. Equations \eref{sbconteom1}-\eref{sbconteom2} can then be used to find solutions for terms in the expansion of the reduced propagator. 

It remains to show that the stationary path approximation yields the same exact result in this example. The propagator for finite time in this case is
\begin{eqnarray}
\fl K(t,0) =\int \prod_{j=1}^{N-1} {\rm d}\mu (\n_j) {\rm d}\mu (z_j) \langle \nb_N \zb_N | \exp(-\frac{{\rm i}}{\hbar} H \ep) | \n_{N-1} z_{N-1} \rangle \nonumber\\
\times \langle \nb_{N-1} \zb_{N-1} | \exp(-\frac{{\rm i}}{\hbar} H \ep) | \n_{N-2} z_{N-2} \rangle ... \langle \nb_{1} \zb_1 | \exp(-\frac{{\rm i}}{\hbar} H \ep) | \n_{0} z_0 \rangle.
\end{eqnarray}
As for the previous example the infinitesmal propagators in the above expression can be combined into a single exponential only after the introduction of a Grassmann partners to the coupling constants. The propagator is then written,
\begin{eqnarray}
\fl K(t,0)=\int \prod_{j=1}^{N-1} {\rm d}^2 \n_j {\rm d}^2 z_j \mbox{ } \exp(\zb_N z_N +\nb_N \n_N) \exp \Bigg \{ \sum_{j=1}^N \bigg[-\zb_j z_j -\nb_j \n_j \nonumber\\
+ (1-{\rm i} \w \ep) \zb_j \z_{j-1} + (1-{\rm i} \w_o \ep) \nb_j \n_{j-1} -\nb_j ({\rm i} \la_j \ep) z_{j-1} -\zb_j ({\rm i} \la_j \ep) \n_{j-1} \bigg] \Bigg \}.
\end{eqnarray}
Varying discretely about the saddle point, equations for the stationary path are found to be
\begin{eqnarray} 
\label{sb2eom1} z_j = (1-{\rm i} \w \ep) \z_{j-1} -{\rm i} \la_j \ep \n_{j-1} \\ 
\label{sb2eom2} \n_j = (1-{\rm i} \w_o \ep) \n_{j-1} -{\rm i} \la_j \ep z_{j-1}
\end{eqnarray}
and the propagator after inserting the correct boundary conditions $\nb_N=\nb_f$, $\n_0=\n_i$, $\zb_N=\zb_f$, and $\z_0=\z_i$  is
\begin{equation}
K(t,0)=\exp(\nb_f \n_N + \zb_f \z_N).
\end{equation}
with the variables $\n_N$ and $\phi_N$ defined by \eref{sb2eom1}-\eref{sb2eom2}. This again is the same as the exact result, thereby demonstrating that the stationary path approximation is exact for the Jaynes-Cummings Hamiltonian. This specific example was computed with the stationary path approximation in \cite{anastopoulos} using a Grassmannian path integral and in \cite{alscher2} using the SU(2) representation. The results here agree with those found in \cite{alscher2}, where it was also found that the stationary path approximation yielded exact results. The range of validity for the Grassmannian path integral method in  \cite{anastopoulos} was resticted to an initial bosonic vaccum state, but for that restricted range they also found the stationary path approximation to be exact.

\section{Discussion}

In this paper the path integral formulation of a spin$-\frac{1}{2}$ propagator in the Grassmann representation was modified by the introduction of time-indexed anticommuting partners to the Grassmann variables. These anticommuting partners make the formulation more consistent by carrying the effect of the nilpotency and anticommutation of the Grassmann variables in the infinitesmal propagator into the full propagator. That is, when the Grassmann variables at intermediate times are integrated over all values, the effect of their anticommutivity will be lost unless anticommuting partners were introduced which carry that effect. The resulting propagator then looks similar to a usual boson coherent state propagator. The difference is that the anticommutivity of the Grassmann partners must be respected during explicit calculation of the propagator.

The introduction of the Grassmann partners in the propagator gives three main advantages. First it allows one to write the propagator as a path integral and thereby reconnect intuitive notions of integration over all paths with actual computation in the Grassmann representation. The concept of the stationary path can then be applied with greater meaning, becoming a specific sequence of mixed c-number/Grassmann values. The range of applicability of the stationary path approximation in the Grassmann representation then widens to include non-vacuum initial states, as shown by the second example. 

The second advantage is that it creates a novel expansion of the propagator as, for example, in \eref{sbprop}. The expansion in \eref{sbprop} turns out to be an expansion in the final coherent state. So for small $| \z_f |$ (small energy) the first few terms are a good approximation. Alternatively, in the same example, $K(t,0)$ can also be written as an expansion in $| \z_i |$ such that the expansion is one near initial vacuum. In that regime the Grassmann representation can be used to find the full quantum dynamics of a spin$-\frac{1}{2}$ (or two-level) system. An example of such a situation would for an atom in a cavity which is near vacuum. As a model, the Hamiltonian of the Jaynes-Cummings example could be extended to include the quantum vacuum, with one mode populated by a few photons. This is the focus of ongoing work.

The third advantage is that it facilitates finding the reduced dynamics since the propagator is in an exponential form. Inserting an initial state and integrating out certain degrees of freedom then requires only a few gaussian integrations. The proviso is that the final result must be expanded in a polynomial series with the anticommutivity of the Grassmann partners respected. This would be especially suited to finding the evolution, including decoherence, of a spin$-\frac{1}{2}$ or two-level system interacting with a bath in some arbitrary initial state. Such an evolution would be non-Markovian, although reduced, since no Markovian approximation is in principle necessary.

\appendix
\section{Spin-$\frac{1}{2}$ in a magnetic field details}
First demonstrate that that the propagator \eref{bprop} with the equations of motion \eref{bconteom1}-\eref{bconteom2} satisfies the Schroedinger equation. The Schroedinger equation is
\begin{equation} \label{sch1}
{\rm i} \hbar \frac{{\rm d}}{{\rm d}t} K(t,0) = H K(t,0).
\end{equation}
In the Grassmannian representation the left hand side(LHS) is from \eref{bprop}
\begin{equation}
{\rm i} \hbar \frac{{\rm d}}{{\rm d}t} \langle \nb_f | K(t,0)  | \n_0 \rangle = {\rm i} \hbar \frac{{\rm d}}{{\rm d}t}\sum_{m=0}^\infty \frac{1}{m!} ([\phi^m]_t+ \nb_f [\phi^m \n]_t).
\end{equation}
Substituting \eref{bconteom1}-\eref{bconteom2} this becomes
\begin{equation}
LHS = \hbar \sum_{m=0}^\infty \frac{1}{m!} (B^* [\phi^m \n]_t +\nb_f B [\phi^m]_t + \w \nb_f [\phi^m \n]_t).
\end{equation}
The right hand side(RHS) of \eref{sch1} is, in the Grassmannian representation
\begin{equation}
\langle \nb_f | H K(t,0) | \n_0 \rangle = \int {\rm d}\mu(\n) \langle \nb_f | H | \n \rangle \langle \nb | K(t,0) | \n_0 \rangle.
\end{equation}
After substituting the Hamiltonian and the propagator \eref{sbprop} the right hand side is
\begin{equation}
RHS = \hbar \sum_{m=0}^\infty \frac{1}{m!} (B^* [\phi^m \n]_t +\nb_f B [\phi^m]_t + \w \nb_f [\phi^m \n]_t).
\end{equation}
Thus in \eref{sch1} LHS=RHS so the Schroedinger equation is satisfied.

\section{Spin-boson details}
The Schroedinger equation is \eref{sch1}. For the spin-boson propagator \eref{sbprop} in the Grassmannian representation the left hand side is
\begin{equation}
{\rm i} \hbar \frac{{\rm d}}{{\rm d}t} \langle \nb_f \zb_f | K(t,0)  | \n_0 \z_0 \rangle = {\rm i} \hbar \frac{{\rm d}}{{\rm d}t}\sum_{m=0}^\infty \frac{(\zb_f)^m}{m!} ( [\z^m]_t +\nb_f [\z^{m}\n]_t).
\end{equation}
Substituting the equations of motion in the spin-boson example \eref{sbconteom1}-\eref{sbconteom2} gives
\begin{equation}
\fl LHS = \hbar \sum_{m=0}^\infty \frac{(\zb_f)^m}{m!} \{ (m\w [\z^m]_t +m \la [\z^{m-1}\n]_t)+ \nb_f ( (m\w +\w_o)[\z^{m}\n]_t + \la [\z^{m+1}]_t ) \}.
\end{equation}
The right hand side of the propagator for a system with both a Grassmannian and a bosonic coherent representation is
\begin{equation}
\langle \nb_f \zb_f | H K(t,0) | \n_0 \z_o \rangle = \int {\rm d}\mu(\n) {\rm d}\mu(\z) \langle \nb_f \zb_f | H | \n \z \rangle \langle \nb \zb | K(t,0) | \n_0 \z_0 \rangle.
\end{equation}
Inserting the Jaynes-Cummings Hamiltonian and evaluating the integral gives
\begin{equation}
\fl RHS = \hbar \sum_{m=0}^\infty \frac{(\zb_f)^m}{m!} \{ (m\w [\z^m]_t +m \la [\z^{m-1}\n]_t)+ \nb_f ( (m\w +\w_o)[\z^{m}\n]_t + \la [\z^{m+1}]_t ) \}
\end{equation}
By inpection LHS=RHS so the Schroedinger equation is satisfied in the spin-boson case.


\section*{References}
\begin{thebibliography}{9}

\bibitem{alscher1}
Alscher A and Grabert H 1999 Semi-classical dynamics of a spin-$\frac{1}{2}$ in an arbitrary magnetic field {\it J. Phys. A:Math. Gen.} {\bf 32} 4907 \\
(Alscher A and Grabert H 1999 {\it Preprint} quant-ph/9904102)

\bibitem{alscher2}
Alscher A and Grabert H 2001 Semi-classical dynamics of the Jaynes-Cummings model {\it Preprint} quant-ph/0006072

\bibitem{boudjeda}
Boudjeda T, Hammann T F and Nouicer Kh 1995 Path integral for a spinning particle in a magnetic field {\it J. Math. Phys.} {\bf 36 4} 1602

\bibitem{smirnov}
Smirnov V V 1999 Path integral for system with spin {\it J. Phys. A:Math. Gen.} {\bf 32} 1285

\bibitem{anastopoulos}
Anastopoulos C and Hu B L 2000 Two-level atom-field interaction: Exact master equation for non-Markovian dynamics, decoherence, and relaxation {\it Phys. Rev. A} {\bf 62} 033821 \\
(Anastopoulos C and Hu B L 1999 {\it Preprint} quant-ph/9901078)

\bibitem{ohnuki}
Ohnuki Y and Kashiwa T 1978 Coherent States of fermi operators and the path integral {\it Coherent states: applications in physics and mathematical physics} eds J Klauder and B Skagerstam (Singapore: World Scientific) pp 449-465

\bibitem{cahill}
Cahill K E and Glauber R J 1999 Density operators for fermions {\it Phys. Rev. A} {\bf 59 2} 1538 \\
(Cahill K E and Glauber R J 1998 {\it Preprint} physics/9808029)

\bibitem{perelomov}
Perelomov A 1986 {\it Generalized coherent states and their applications} (Berlin: Springer)

\end{thebibliography}
\end{document}